\documentstyle[sprocl]{article}
\input{psfig}
\def\Journal#1#2#3#4{{#1} {\bf #2}, #3 (#4)}

\def\PLB{{\em Phys. Lett.}  B}

\def\PRD{{\em Phys. Rev.} D}

\def\be{\begin{equation}}
\def\ee{\end{equation}}
\def\bea{\begin{eqnarray}}
\def\eea{\end{eqnarray}}
\begin{document}
\vspace*{-2.5cm}
\begin{flushright}
UR-1477 \\
ER-40685-893 \\
hep-ph/9609354 \\
     September 1996 \\
\end{flushright}
\vskip .75cm

\title{GLUON RADIATION AND TOP QUARK PHYSICS\footnote{Presented 
by L.H.\ Orr at the 1996 Meeting of the Division of Particles and 
Fields, Minneapolis, MN, August 10--15, 1996}}
\author{ L.H. ORR }
\address{Department of Physics and Astronomy, University of Rochester, \\
Rochester, NY 14627, USA}
\author{ T. STELZER }
\address{Department of Physics, University of Illinois-Urbana \\
Urbana, Illinois 61801, USA}
\author{ W.J. STIRLING }
\address{Departments of Physics and Mathematical Sciences, 
University of Durham \\
Durham DH1 3LE, England }
\maketitle\abstracts{
Radiation of gluons gives rise to extra jets in top quark events that
can lead to complications in event reconstruction and mass measurement.
I review recent results for gluon radiation in top quark production and
decay, and discuss their implications for top quark physics.}

\section{Tevatron}
The presence of extra jets in $t \bar t$ events can complicate attempts to 
reconstruct the mass of the top quark from the momenta of its decay products.
Depending on whether the associated gluon was emitted during top 
production or decay, it may or may not be appropriate to include 
its momentum in top reconstruction.  Unfortunately, it is not easy to
tell the difference.  Therefore we must consider top production and decay
together in our treatment of gluon radiation.   

We have computed the complete order $\alpha_s^3$ tree-level 
cross section for $t \bar t$ production and 
decay with an extra jet at the Tevatron~\cite{OSS} and the LHC~\cite{OSSLHC};
for more details see the references.
At the Tevatron we find~\cite{OSS} that the contribution (after kinematic cuts) 
from gluons radiated in the top production process is comparable to 
that from decay.  As we might expect, the decay-stage gluons appear 
mostly in the central rapidity region.  Production-stage radiation occupies
a wider rapidity range, but it, too, contributes substantially in the 
central region.  The distribution in the angle between the extra jet and 
the $b$ quark is dominated at small angles by the decay-stage contribution,
but the production-stage piece is non-negligible in that region as
well.  
The bottom line is that there is no clear way experimentally to 
distinguish the two contributions, and they must be taken into account
together in our analyses.

\section{LHC}

\begin{figure}
\vspace{14cm}
\vspace{6.5cm}
\hspace{-4.5cm}
\includegraphics{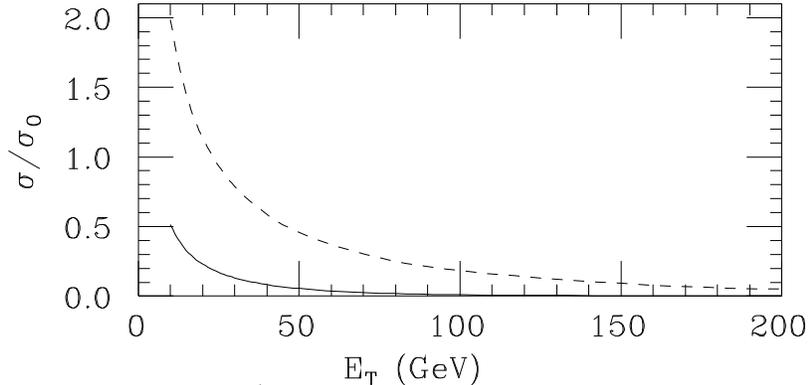}
\vspace{-16.4cm}
\caption{$\sigma(t\bar t j)/\sigma(t \bar t)$ for $E_T^j >E_T$
at the Tevatron (solid line) and LHC (dashed line).
\label{fig:sigmas}}
\end{figure}

The higher energy of the LHC compared to the Tevatron leads to a large
increase in radiation in $t \bar t$ events, most of it associated with
production~\cite{OSSLHC}.  
The increase is due to the larger color charge of the 
$gg$ initial state which dominates at the LHC, the behavior of the 
parton distributions, and the phase space available for radiating a gluon.
The amounts of production-stage radiation at the Tevatron and LHC can
be seen in Fig.~\ref{fig:sigmas} where we show 
$\sigma(t\bar t j, E_{Tj}>E_T)/\sigma(t\bar t)$ as a function of $E_T$.
For all values of the $E_T$ cut, the ratio of (tree-level) NLO to LO 
cross sections is much larger at the LHC.

The rapidity distribution of extra jets in top production and decay 
at the LHC is relatively flat.  It is dominated by the production-stage
contribution; the exact proportions depend on cuts on the jet $E_T$ 
and $\Delta R$ between the jet and $b$ quarks.  The distribution in 
$\Delta R$ is dominated by the production contribution for all values above
0.4.  This abundance of radiation at the LHC can lead to difficulties
in momentum reconstruction when, for example, production stage
radiation gets included in top decay products' jet cones.

\begin{figure}[eett]
\psfig{figure=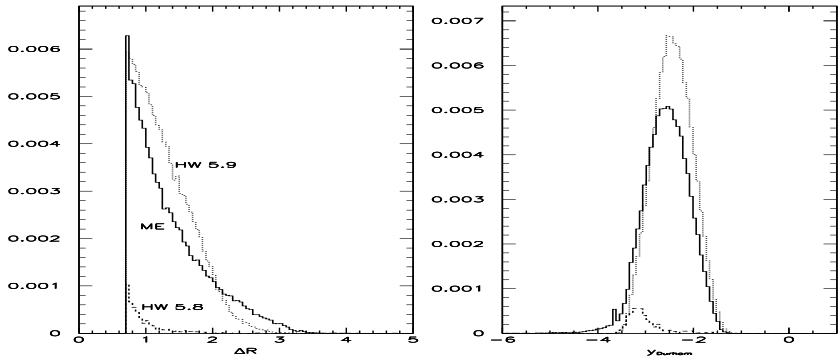,height=1.9in,width=\textwidth}
\caption{Distributions in (a) the minimum 
$\Delta R$ and (b) the minimum $y$ 
for additional jets produced in 
$e^+e^-$ collisions at 360 GeV c.m.\ energy.
\label{fig:hwg}}
\end{figure}

\section{Comparison with HERWIG}

It is useful to compare the results from the fixed-order QCD 
calculation described above to those obtained 
from commonly used Monte Carlo programs 
which generate extra jets via parton showers.
In previous work,~\cite{OSS,OSSBIS} a comparison of our results with those from
HERWIG version 5.8~\cite{HERWIG} revealed discrepancies
in regions where the two calculations  should agree.  
HERWIG 5.8 was later shown to contain a bug which caused decay-stage radiation
to be suppressed.

In the most recent version of HERWIG (5.9)~\cite{HERWIG}, the 
restoration of decay-stage radiation leads to better agreement.  Some
questions still remain, however:~\cite{OSSLHC}  there now appears to be
too {\it much} decay-stage radiation in HERWIG compared to the matrix-element 
result.  This is most easily seen for $e^+e^-$ colliders, where there 
are no complications due to initial-state gluon radiation.  We consider
gluon radiation in top production and decay at center-of-mass energy 360 GeV,
just above top pair threshold, so that most of the gluons originate in the 
decays.  In Fig.~\ref{fig:hwg} we show the distributions in the minimum 
$\Delta R$  and $y_{\rm Durham}$ for jet pairs in top events.
We see that HERWIG 5.8, with
the bug, vastly underestimates the exact matrix-element distribution,
while HERWIG 5.9 overestimates it.
More subtle is an additional discrepancy in the production-stage radiation
which appears on the jet rapidity distribution at hadron colliders.
The HERWIG distribution shows a slight dip in the central region; the 
matrix element distribution does not.  It appears that HERWIG still needs
a hard correction for gluon radiation from quarks as heavy as top. 

As statistics improve, systematic uncertainties
associated with gluon radiation will dominate top physics, 
$m_t$ measurements especially.  It is important that the relevant
physics is incorporated correctly in experimental analyses.

\section*{Acknowledgments} 
This work was supported in part by U.S.\
D.O.E.\ grant DE-FG02-91ER40685.


\section*{References}
\begin{thebibliography}{99} 
\bibitem{OSS}L.H.~Orr, T.~Stelzer and W.J.~Stirling,
\Journal{\PRD}{52}{124}{1995}.
\bibitem{OSSLHC}L.H.~Orr, T.~Stelzer and W.J.~Stirling, preprint UR-1473,
August 1996.
\bibitem{OSSBIS}L.H.~Orr, T.~Stelzer and W.J.~Stirling,
\Journal{\PLB}{354}{442}{1995}.
\bibitem{HERWIG} G.~Marchesini {\it et al.,}
hep-ph/9607393, and references therein.
\end{thebibliography}
\end{document}